\documentclass[12pt]{article}
\usepackage{amssymb,amscd,array}
\catcode `\@=11
\@addtoreset{equation}{section}

\newtheorem{thm}{Theorem}[section]

\newtheorem{cor}[thm]{Corollary}

\def\qed{\blacksquare}
\newcommand{\be}{\begin{equation}}
\newcommand{\ee}{\end{equation}}
\newcommand{\bea}{\begin{eqnarray}}
\newcommand{\eea}{\end{eqnarray}}
\newcommand{\R}{\mathbb{R}}
\newcommand{\N}{\mathbb{N}}
\newcommand{\C}{\mathbb{C}}

\begin{document}
\begin{titlepage}

\begin{center}
{\bf \Large{On the Super-Renormalizablity of Gauge Models in the Causal
Approach\\}}
\end{center}
\vskip 1.0truecm
\centerline{D. R. Grigore, 
\footnote{e-mail: grigore@theory.nipne.ro}}
\vskip5mm
\centerline{Department of Theoretical Physics}
\centerline{Institute for Physics and Nuclear Engineering ``Horia Hulubei"}
\centerline{Bucharest-M\u agurele, P. O. Box MG 6, ROM\^ANIA}

\vskip 2cm
\bigskip \nopagebreak
\begin{abstract}
\noindent
We consider some typical gauge models in the causal approach: Yang-Mills and
pure 
massless gravity up to the second order of the perturbation theory. We prove
that
the loop contributions are coboundaries, up to super-renormalizable terms in
the 
Yang-Mills case; this means that the ultra-violet behavior is better than
expected
from power counting considerations. For the pure massless gravity we prove that
the 
loop contributions are coboundaries so the model is essentially classical. We 
conjecture that such a result should be true in higher orders of the
perturbation
theory also. This result should make easier the problem of constructive quantum
field
theory.
\end{abstract}
\end{titlepage}

\section{Introduction}

The general framework of perturbation theory consists in the construction of 
the chronological products: for every set of Wick monomials 
$ 
W_{1}(x_{1}),\dots,W_{n}(x_{n}) 
$
acting in some Fock space
$
{\cal H}
$
one associates the operator
$ 
T^{W_{1},\dots,W_{n}}(x_{1},\dots,x_{n}); 
$  
all these expressions are in fact distribution-valued operators called
chronological products. It will be convenient to use another notation: 
$ 
T(W_{1}(x_{1}),\dots,W_{n}(x_{n})). 
$ 
These operators are constrained by Bogoliubov axioms  \cite{BS}, \cite{EG},
\cite{DF}; we prefer the setting from \cite{DF}. (An equivalent
point of view uses retarded products \cite{St1}.) The construction of
the chronological products can be done recursively according to Epstein-Glaser
prescription \cite{EG}, \cite{Gl} (which reduces the induction procedure to a
distribution splitting of some distributions with causal support) or according
to Stora prescription \cite{PS} (which reduces the renormalization procedure to
the process of extension of distributions). These products are not uniquely
defined but there are some natural limitation on the arbitrariness. If this
arbitrariness does not grow with the order $n$ of the perturbation theory then
we say that the theory is renormalizable; the most popular point of view is that
only such theories are physically meaningful.

Gauge theories describe particles of higher spin. Usually such theories are not
renormalizable. However, one can save renormalizability using ghost fields.
Such theories are defined in a Fock space
$
{\cal H}
$
with indefinite metric, generated by physical and un-physical fields (called
{\it ghost fields}). One selects the physical states assuming the existence of
an operator $Q$ called {\it gauge charge} which verifies
$
Q^{2} = 0
$
and such that the {\it physical Hilbert space} is by definition
$
{\cal H}_{\rm phys} \equiv Ker(Q)/Im(Q).
$
The space
$
{\cal H}
$
is endowed with a grading (usually called {\it ghost number}) and by
construction the gauge charge is raising the ghost number of a state. Moreover,
the space of Wick monomials in
$
{\cal H}
$
is also endowed with a grading which follows by assigning a ghost number to
every one of the free fields generating
$
{\cal H}.
$
The graded commutator
$
d_{Q}
$
of the gauge charge with any operator $A$ of fixed ghost number
\be
d_{Q}A = [Q,A]
\ee
is raising the ghost number by a unit. Because
\be
d_{Q}^{2} = 0
\ee
it means that
$
d_{Q}
$
is a co-chain operator in the space of Wick polynomials. From now on
$
[\cdot,\cdot]
$
denotes the graded commutator.
 
A gauge theory assumes also that there exists a Wick polynomial of null ghost
number
$
T(x)
$
called {\it the interaction Lagrangian} such that
\be
~[Q, T] = i \partial_{\mu}T^{\mu}
\label{gau1}
\ee
for some other Wick polynomials
$
T^{\mu}.
$
This relation means that the expression $T$ leaves invariant the physical
states, at least in the adiabatic limit. Indeed, if this is true we have:
\be
T(f)~{\cal H}_{\rm phys}~\subset~~{\cal H}_{\rm phys}  
\label{gau2}
\ee
up to terms which can be made as small as desired (making the test function $f$
flatter and flatter). In all known models one finds out that there exist a chain
of Wick polynomials
$
T^{\mu},~T^{\mu\nu},~T^{\mu\nu\rho},\dots
$
such that:
\be
~[Q, T] = i \partial_{\mu}T^{\mu}, \quad
[Q, T^{\mu}] = i \partial_{\nu}T^{\mu\nu}, \quad
[Q, T^{\mu\nu}] = i \partial_{\rho}T^{\mu\nu\rho},\dots
\label{descent}
\ee
It so happens that for all these models the expressions
$
T^{\mu\nu},~T^{\mu\nu\rho},\dots
$
are completely antisymmetric in all indexes; it follows that the chain of
relation stops at the step $4$ (if we work in four dimensions). We can also use
a compact notation
$
T^{I}
$
where $I$ is a collection of indexes
$
I = [\nu_{1},\dots,\nu_{p}]~(p = 0,1,\dots,)
$
and the brackets emphasize the complete antisymmetry in these indexes. All these
polynomials have the same canonical dimension
\be
\omega(T^{I}) = \omega_{0},~\forall I
\ee
and because the ghost number of
$
T \equiv T^{\emptyset}
$
is supposed null, then we also have:
\be
gh(T^{I}) = |I|.
\ee
One can write compactly the relations (\ref{descent}) as follows:
\be
d_{Q}T^{I} = i~\partial_{\mu}T^{I\mu}.
\label{descent1}
\ee
For concrete models the equations (\ref{descent}) can stop earlier: for 
instance in the Yang-Mills case we have
$
T^{\mu\nu\rho} = 0
$
and in the case of gravity
$
T^{\mu\nu\rho\sigma} = 0.
$
If the interaction Lagrangian $T$ is Lorentz invariant, then one can prove that
the expressions
$
T^{I},~|I| > 0
$
can be taken Lorentz covariant.

Now we can construct the chronological products
$$
T^{I_{1},\dots,I_{n}}(x_{1},\dots,x_{n}) \equiv
T(T^{I_{1}}(x_{1}),\dots,T^{I_{n}}(x_{n}))
$$
according to the recursive procedure. We say that the theory is gauge invariant
in all orders of the perturbation theory if the following set of identities
generalizing (\ref{descent1}):
\be
d_{Q}T^{I_{1},\dots,I_{n}} = 
i \sum_{l=1}^{n} (-1)^{s_{l}} {\partial\over \partial x^{\mu}_{l}}
T^{I_{1},\dots,I_{l}\mu,\dots,I_{n}}
\label{gauge}
\ee
are true for all 
$n \in \N$
and all
$
I_{1}, \dots, I_{n}.
$
Here we have defined
\be
s_{l} \equiv \sum_{j=1}^{l-1} |I|_{j}.
\ee

We introduce some cohomology terminology. We consider a {\it cochains} to be
an ensemble of distribution-valued operators of the form
$
C^{I_{1},\dots,I_{n}}(x_{1},\dots,x_{n}),~n = 1,2,\cdots
$
(usually we impose some supplementary symmetry properties) and define the
derivative operator $\delta$ according to
\be
(\delta C)^{I_{1},\dots,I_{n}}
= \sum_{l=1}^{n} (-1)^{s_{l}} {\partial\over \partial x^{\mu}_{l}}
C^{I_{1},\dots,I_{l}\mu,\dots,I_{n}}.
\ee
We can prove that 
\be
\delta^{2} = 0.
\ee
Next we define
\be
s = d_{Q} - i \delta,\qquad \bar{s} = d_{Q} + i \delta
\ee
and note that
\be
s \bar{s} = \bar{s} s = 0.
\ee
We call {\it relative cocycles} the expressions $C$ verifying
\be
sC = 0
\ee
and a {\it relative coboundary} an expression $C$ of the form
\be
C = \bar{s}B.
\ee
The relation (\ref{gauge}) is simply the cocycle condition
\be
sT = 0.
\ee

The purpose of this paper is to investigate if this condition implies that, at
least some contributions of $T$, are in fact coboundaries. Coboundaries are 
trivial from the physical point of view: if we consider two physical states
$
\Psi, \Psi^{\prime}
$
then
\be
<\Psi, \bar{s}B\Psi^{\prime}> = 
< Q\Psi, B\Psi^{\prime}> - < \Psi, BQ\Psi^{\prime}>
+ i < \Psi, \delta B\Psi^{\prime}> \rightarrow 0
\ee
(in the adiabatic limit). 

We will consider here only the second order of the perturbation theory and prove
that for Yang-Mills models the loop contributions are coboundaries, up to
super-renormalizable terms (i.e. terms with a better ultra-violet behavior than
given by power counting); for massless gravity the situation is even better,
i.e. the loop contributions are strictly a coboundary i.e. the theory is
essentially classical. This follows from the fact that in the loop expansion
the $0$-loop (or tree) contribution corresponds to the classical theory
\cite{DF3}.

In the next Section we present the description of the free fields use, mainly to
fix the notations. In Section \ref{ggt} we give Bogoliubov axioms for the second
order of the perturbation theory; in Subsection \ref{causal} we give the basic
distributions with causal support appearing for loop contributions in the second
order of the perturbation theory. In Sections \ref{ym} and \ref{gravity} we
prove
the cohomology result for Yang-Mills and gravity.
\newpage

\section{Free Fields\label{free}}
We summarize some results from \cite{cohomology}.
\subsection{Massless Particles of Spin $1$ (Photons)}

We consider a vector space 
$
{\cal H}
$
of Fock type generated (in the sense of Borchers theorem) by the vector field 
$
v_{\mu}
$ 
(with Bose statistics) and the scalar fields 
$
u, \tilde{u}
$
(with Fermi statistics). The Fermi fields are usually called {\it ghost fields}.
We suppose that all these (quantum) fields are of null mass. Let $\Omega$ be the
vacuum state in
$
{\cal H}.
$
In this vector space we can define a sesquilinear form 
$<\cdot,\cdot>$
in the following way: the (non-zero) $2$-point functions are by definition:
\bea
<\Omega, v_{\mu}(x_{1}) v_{\mu}(x_{2})\Omega> =i~\eta_{\mu\nu}~D_{0}^{(+)}(x_{1}
- x_{2}),
\nonumber \\
<\Omega, u(x_{1}) \tilde{u}(x_{2})\Omega> =- i~D_{0}^{(+)}(x_{1} - x_{2})
\qquad
<\Omega, \tilde{u}(x_{1}) u(x_{2})\Omega> = i~D_{0}^{(+)}(x_{1} - x_{2})
\eea
and the $n$-point functions are generated according to Wick theorem. Here
$
\eta_{\mu\nu}
$
is the Minkowski metrics (with diagonal $1, -1, -1, -1$) and 
$
D_{0}^{(+)}
$
is the positive frequency part of the Pauli-Jordan distribution
$
D_{0}
$
of null mass. To extend the sesquilinear form to
$
{\cal H}
$
we define the conjugation by
\be
v_{\mu}^{\dagger} = v_{\mu}, \qquad 
u^{\dagger} = u, \qquad
\tilde{u}^{\dagger} = - \tilde{u}.
\ee

Now we can define in 
$
{\cal H}
$
the operator $Q$ according to the following formulas:
\bea
~[Q, v_{\mu}] = i~\partial_{\mu}u,\qquad
[Q, u] = 0,\qquad
[Q, \tilde{u}] = - i~\partial_{\mu}v^{\mu}
\nonumber \\
Q\Omega = 0
\label{Q-0}
\eea
where by 
$
[\cdot,\cdot]
$
we mean the graded commutator. One can prove that $Q$ is well defined. Indeed,
we have the causal commutation relations 
\be
~[v_{\mu}(x_{1}), v_{\mu}(x_{2}) ] =i~\eta_{\mu\nu}~D_{0}(x_{1} - x_{2})~\cdot
I,
\qquad
[u(x_{1}), \tilde{u}(x_{2})] = - i~D_{0}(x_{1} - x_{2})~\cdot I
\ee
and the other commutators are null. The operator $Q$ should leave invariant
these relations, in particular 
\be
[Q, [ v_{\mu}(x_{1}),\tilde{u}(x_{2})]] + {\rm cyclic~permutations} = 0
\ee
which is true according to (\ref{Q-0}). It is useful to introduce a grading in 
$
{\cal H}
$
as follows: every state which is generated by an even (odd) number of ghost
fields and an arbitrary number of vector fields is even (resp. odd). We denote
by 
$
|f|
$
the ghost number of the state $f$. We notice that the operator $Q$ raises the
ghost number of a state (of fixed ghost number) by an unit. The usefulness of
this construction follows from:
\begin{thm}
The operator $Q$ verifies
$
Q^{2} = 0.
$ 
The factor space
$
Ker(Q)/Ran(Q)
$
is isomorphic to the Fock space of particles of zero mass and helicity $1$
(photons). 
\end{thm}

\newpage

\subsection{Massive Particles of Spin $1$ (Heavy Bosons)}

We repeat the whole argument for the case of massive photons i.e. particles of
spin $1$ and positive mass. 

We consider a vector space 
$
{\cal H}
$
of Fock type generated (in the sense of Borchers theorem) by the vector field 
$
v_{\mu},
$ 
the scalar field 
$
\Phi
$
(with Bose statistics) and the scalar fields 
$
u, \tilde{u}
$
(with Fermi statistics). We suppose that all these (quantum) fields are of mass
$
m > 0.
$
In this vector space we can define a sesquilinear form 
$<\cdot,\cdot>$
in the following way: the (non-zero) $2$-point functions are by definition:
\bea
<\Omega, v_{\mu}(x_{1}) v_{\mu}(x_{2})\Omega> =i~\eta_{\mu\nu}~D_{m}^{(+)}(x_{1}
- x_{2}),
\quad
<\Omega, \Phi(x_{1}) \Phi(x_{2})\Omega> =- i~D_{m}^{(+)}(x_{1} - x_{2})
\nonumber \\
<\Omega, u(x_{1}) \tilde{u}(x_{2})\Omega> =- i~D_{m}^{(+)}(x_{1} - x_{2}),
\qquad
<\Omega, \tilde{u}(x_{1}) u(x_{2})\Omega> = i~D_{m}^{(+)}(x_{1} - x_{2})
\eea
and the $n$-point functions are generated according to Wick theorem. Here
$
D_{m}^{(+)}
$
is the positive frequency part of the Pauli-Jordan distribution
$
D_{m}
$
of mass $m$. To extend the sesquilinear form to
$
{\cal H}
$
we define the conjugation by
\be
v_{\mu}^{\dagger} = v_{\mu}, \qquad 
u^{\dagger} = u, \qquad
\tilde{u}^{\dagger} = - \tilde{u},
\qquad \Phi^{\dagger} = \Phi.
\ee

Now we can define in 
$
{\cal H}
$
the operator $Q$ according to the following formulas:
\bea
~[Q, v_{\mu}] = i~\partial_{\mu}u,\qquad
[Q, u] = 0,\qquad
[Q, \tilde{u}] = - i~(\partial_{\mu}v^{\mu} + m~\Phi)
\qquad
[Q,\Phi] = i~m~u,
\nonumber \\
Q\Omega = 0.
\label{Q-m}
\eea
One can prove that $Q$ is well defined. We have a result similar to the first
theorem of this Section:
\begin{thm}
The operator $Q$ verifies
$
Q^{2} = 0.
$ 
The factor space
$
Ker(Q)/Ran(Q)
$
is isomorphic to the Fock space of particles of mass $m$ and spin $1$ (massive
photons). 
\end{thm}

\newpage
\subsection{The Generic Yang-Mills Case\label{generic}}

The situations described above (of massless and massive photons) are susceptible
of the following generalizations. We can consider a system of 
$
r_{1}
$ 
species of particles of null mass and helicity $1$ if we use in the first part
of this Section 
$
r_{1}
$ 
triplets
$
(v^{\mu}_{a}, u_{a}, \tilde{u}_{a}), a \in I_{1}
$
of massless fields; here
$
I_{1}
$
is a set of indices of cardinal 
$
r_{1}.
$
All the relations have to be modified by appending an index $a$ to all these
fields. 

In the massive case we have to consider 
$
r_{2}
$ 
quadruples
$
(v^{\mu}_{a}, u_{a}, \tilde{u}_{a}, \Phi_{a}),  a \in I_{2}
$
of fields of mass 
$
m_{a}
$; here
$
I_{2}
$
is a set of indexes of cardinal 
$
r_{2}.
$

We can consider now the most general case involving fields of spin not greater
that $1$.
We take 
$
I = I_{1} \cup I_{2} \cup I_{3}
$
a set of indexes and for any index we take a quadruple
$
(v^{\mu}_{a}, u_{a}, \tilde{u}_{a},\Phi_{a}), a \in I
$
of fields with the following conventions:
(a) For
$
a \in I_{1}
$
we impose 
$
\Phi_{a} = 0
$
and we take the masses to be null
$
m_{a} = 0;
$
(b) For
$
a \in I_{2}
$
we take the all the masses strictly positive:
$
m_{a} > 0;
$
(c) For 
$
a \in I_{3}
$
we take 
$
v_{a}^{\mu}, u_{a}, \tilde{u}_{a}
$
to be null and the fields
$
\Phi_{a} \equiv \phi^{H}_{a} 
$
of mass 
$
m^{H}_{a} \geq 0.
$
The fields
$
\phi^{H}_{a} 
$
are called {\it Higgs fields}.

If we define
$
m_{a} = 0, \forall a \in I_{3}
$
then we can define in 
$
{\cal H}
$
the operator $Q$ according to the following formulas for all indexes
$
a \in I:
$
\bea
~[Q, v^{\mu}_{a}] = i~\partial^{\mu}u_{a},\qquad
[Q, u_{a}] = 0,
\nonumber \\
~[Q, \tilde{u}_{a}] = - i~(\partial_{\mu}v^{\mu}_{a} + m_{a}~\Phi_{a})
\qquad
[Q,\Phi_{a}] = i~m_{a}~u_{a},
\nonumber \\
Q\Omega = 0.
\label{Q-general}
\eea

If we consider matter fields also i.e some set of Dirac fields with Fermi
statistics:
$
\Psi_{A}, A \in I_{4}
$ 
then we impose
\be
d_{Q}\Psi_{A} = 0. 
\ee

\newpage
\subsection{Massless Particles of Spin $2$ (Gravitons)}

We consider the vector space 
$
{\cal H}
$
of Fock type generated (in the sense of Borchers theorem) by the symmetric
tensor field 
$
h_{\mu\nu}
$ 
(with Bose statistics) and the vector fields 
$
u^{\rho}, \tilde{u}^{\sigma}
$
(with Fermi statistics). We suppose that all these (quantum) fields are of null mass. 
In this vector space we can define a sesquilinear form 
$<\cdot,\cdot>$
in the following way: the (non-zero) $2$-point functions are by definition:
\bea
<\Omega, h_{\mu\nu}(x_{1}) h_{\rho\sigma}(x_{2})\Omega> = - {i\over 2}~
(\eta_{\mu\rho}~\eta_{\nu\sigma} + \eta_{\nu\rho}~\eta_{\mu\sigma}
- \eta_{\mu\nu}~\eta_{\rho\sigma})~D_{0}^{(+)}(x_{1} - x_{2}),
\nonumber \\
<\Omega, u_{\mu}(x_{1}) \tilde{u}_{\nu}(x_{2})\Omega> = i~\eta_{\mu\nu}~
D_{0}^{(+)}(x_{1} - x_{2}),
\nonumber \\
<\Omega, \tilde{u}_{\mu}(x_{1}) u_{\nu}(x_{2})\Omega> = - i~\eta_{\mu\nu}~
D_{0}^{(+)}(x_{1} - x_{2})
\eea
and the $n$-point functions are generated according to Wick theorem. Here
$
\eta_{\mu\nu}
$
is the Minkowski metrics (with diagonal $1, -1, -1, -1$) and 
$
D_{0}^{(+)}
$
is the positive frequency part of the Pauli-Jordan distribution
$
D_{0}
$
of null mass. To extend the sesquilinear form to
$
{\cal H}
$
we define the conjugation by
\be
h_{\mu\nu}^{\dagger} = h_{\mu\nu}, \qquad 
u_{\rho}^{\dagger} = u_{\rho}, \qquad
\tilde{u}_{\sigma}^{\dagger} = - \tilde{u}_{\sigma}.
\ee

Now we can define in 
$
{\cal H}
$
the operator $Q$ according to the following formulas:
\bea
~[Q, h_{\mu\nu}] = - {i\over 2}~(\partial_{\mu}u_{\nu} + \partial_{\nu}u_{\mu}
- \eta_{\mu\nu} \partial_{\rho}u^{\rho}),\qquad
[Q, u_{\mu}] = 0,\qquad
[Q, \tilde{u}_{\mu}] = i~\partial^{\nu}h_{\mu\nu}
\nonumber \\
Q\Omega = 0
\label{Q-0-2}
\eea
where by 
$
[\cdot,\cdot]
$
we mean the graded commutator. One can prove that $Q$ is well defined. Indeed,
we have the causal commutation relations 
\bea
~[h_{\mu\nu}(x_{1}), h_{\rho\sigma}(x_{2}) ] = - {i\over 2}~
(\eta_{\mu\rho}~\eta_{\nu\sigma} + \eta_{\nu\rho}~\eta_{\mu\sigma}
- \eta_{\mu\nu}~\eta_{\rho\sigma})~D_{0}(x_{1} - x_{2})~\cdot I,
\nonumber \\
~[u(x_{1}), \tilde{u}(x_{2})] = i~\eta_{\mu\nu}~D_{0}(x_{1} - x_{2})~\cdot I
\eea
and the other commutators are null. The operator $Q$ should leave invariant
these relations, in particular 
\be
[Q, [ h_{\mu\nu}(x_{1}),\tilde{u}_{\sigma}(x_{2})]] + {\rm cyclic~permutations}
= 0
\ee
which is true according to (\ref{Q-0}). Then we have:
\begin{thm}
The operator $Q$ verifies
$
Q^{2} = 0.
$ 
The factor space
$
Ker(Q)/Im(Q)
$
is isomorphic to the Fock space of particles of zero mass and helicity $2$
(gravitons). 
\label{fock-0}
\end{thm}
\newpage
\section{General Gauge Theories\label{ggt}}
 
We give here the essential ingredients of perturbation theory for 
$
n = 2
$.
First we consider that the canonical dimension of the vector and scalar fields
$
v_{a}^{\mu}, u_{a}, \tilde{u}_{a}, \Phi_{a}
$
and
$
h_{\mu\nu}, u_{\mu}, \tilde{u}_{\mu}
$
is equal to $1$ and the canonical dimension of the Dirac fields is
$
3/2
$.
A derivative applied to a field raises the canonical dimension by $1$. The ghost
number of the ghost fields is $1$ and for the rest of the fields is null. The
Fermi parity of a Fermi (Bose) field is $1$ (resp. $0$). The canonical 
dimension of a Wick monomial is additive with respect to the factors and the
same is true for the ghost number and the Fermi parity.
\subsection{Bogoliubov Axioms}{\label{bogoliubov}}

Suppose that the Wick monomials
$
A, B
$
are self-adjoint:
$
A^{\dagger} = A,~B^{\dagger} = B
$
and of fixed Fermi parity 
$
|A|, |B|
$
and canonical dimension
$
\omega(A), \omega(B)
$.
We will consider two case: for Yang-Mills fields we can take
$
\omega(A), \omega(B) \leq 4
$
but for gravity we have
$
\omega(A), \omega(B) \leq 5
$.
The chronological products
$ 
T(A(x),B(y))
$
are verifying the following set of axioms:
\begin{itemize}
\item
Skew-symmetry:
\be
T(B(y),A(x)) = (-1)^{|A||B| } T(A(x),B(y))
\ee
\item
Poincar\'e invariance: we have a natural action of the Poincar\'e group in the
space of Wick monomials and we impose that for all elements $g$ of the
universal 
covering group
$
inSL(2,\C)
$
of the Poincar\'e group:
\be
U_{g} T(A(x),B(y)) U^{-1}_{g} = T(g\cdot A(g\cdot x),g\cdot B(g\cdot y))
\label{invariance}
\ee
where
$
x \mapsto g\cdot x
$
is the action of
$
inSL(2,\C)
$
on the Minkowski space. Sometimes we can supplement this axiom with other
symmetry 
properties, as for instance, parity invariance.
\item
Causality: if
$
x \geq y
$
i.e.
$
(x - y)^{2} \geq 0,~x^{0} - y^{0} \geq 0
$
then we have:
\be
T(A(x),B(y)) = A(x)~B(y);
\label{causality}
\ee
\item
Unitarity: If we define the {\it anti-chronological products} according to
\be
\bar{T}(A(x),B(y)) \equiv A(x) B(y) + B(y) A(x) - T(A(x),B(y)) 
\label{antichrono}
\ee
then the unitarity axiom is:
\be
\bar{T}(A(x),B(y) = T(A(x),B(y))^{\dagger}.
\label{unitarity}
\ee
\end{itemize}

It can be proved that this system of axioms can be supplemented with
\be
T(A(x),B(y)) = \sum \quad
<\Omega, T(A_{1}(x),B_{1}(y))\Omega>~:A_{2}(x)B_{2}(y):
\label{wick-chrono2}
\ee
where
$
A = A_{1}A_{2}, B = B_{1}B_{2}
$
is an arbitrary decomposition of $A$ and resp. $B$ in Wick submonomials and we
have supposed for simplicity that no Fermi fields are present; if Fermi fields 
are present, then some apropriate signs do appear. This is called the {\it Wick
expansion property}. 

We can also include in the induction hypothesis a limitation on the order of
singularity of the vacuum averages of the chronological products:
\be
\omega(<\Omega, T(A(x),B(y))\Omega>) \leq
\omega(A) + \omega(B) - 4
\label{power}
\ee
where by
$\omega(d)$
we mean the order of singularity of the (numerical) distribution $d$ and by
$\omega(W)$
we mean the canonical dimension of the Wick monomial $W$.

The contributions verifying
\be
\omega(<\Omega, T(A(x),B(y))\Omega>) <
\omega(A) + \omega(B) - 4
\label{superpower}
\ee
will be called {\it super-renormalizable}.

The operator-valued distributions
$
D, A, R, T
$
admit a decomposition into loop contributions
$
D = \sum _{l} D_{(l)}
$,
etc. Indeed every contribution is associated with a certain Feynman graph and
the 
integer $l$ counts the number of the loops. Alternatively, if we consider the
loop
decomposition of the advanced (or retarded) products we have in fact series in
$
\hslash
$
so the contribution corresponding to
$
l = 0
$
(the tree contribution) is the classical part and the loop contributions
$
l > 0
$
are the quantum corrections \cite{DF3}.
\subsection{Second Order Cohomology}
We go to the second order of perturbation theory using the {\it causal
commutator}
\be
D^{A,B}(x,y) \equiv D(A(x),B(y)) = [ A(x),B(y)]
\ee
where 
$
A(x), B(y)
$
are arbitrary Wick monomials and, as always we mean by 
$
[\cdot,\cdot]
$
the graded commutator. These type of distributions are translation invariant
i.e. they depend only on 
$
x - y
$
and the support is inside the light cones:
\be
supp(D) \subset V^{+} \cup V^{-}.
\ee

A theorem from distribution theory guarantees that one can causally split this
distribution:
\be
D(A(x),B(y)) = A(A(x),B(y)) - R(A(x),B(y)).
\ee
where:
\be
supp(A) \subset V^{+} \qquad supp(R) \subset V^{-}.
\ee
The expressions 
$
A(A(x),B(y)), R(A(x),B(y))
$
are called {\it advanced} resp. {\it retarded} products. They are not uniquely
defined: one can modify them with {\it quasi-local terms} i.e. terms
proportional with
$
\delta(x - y)
$
and derivatives of it. 

There are some limitations on these redefinitions coming from Lorentz
invariance, and {\it power counting}: this means that we should not make the
various distributions appearing in the advanced and retarded products too
singular.

Then we define the {\it chronological product} by:
\be
T(A(x),B(y)) = A(A(x),B(y)) + B(y) A(x) = R(A(x),B(y)) + A(x) B(y).
\ee

We consider that $A$ and $B$ are of the type 
$
T^{I}
$
such that we have first-order gauge invariance:
\be
sT = 0
\label{gauge1}
\ee
which is a cocycle equation. Then we define that causal commutator
\be
D^{IJ}(x,y) \equiv [T^{I}(x), T^{J}(y)];
\ee
we have the symmetry property
\be
D^{JI}(y,x) = - (-1)^{|I||J|}~D^{IJ}(x,y)
\ee
and the limitations
\be
gh(D^{IJ}) = |I|+ |J|
\ee
and power counting limitations coming from (\ref{power}). This will be our
cochain space.
But
$
D^{IJ}(x,y)
$
it is also a cocycle:
\be
sD = 0 \qquad \Leftrightarrow \qquad
d_{Q}D^{IJ} = i {\partial\over \partial x^{\rho}}D^{I\rho,J} 
+ i (-1)^{|I|} {\partial\over \partial y^{\rho}}D^{I,J\rho}.
\label{cocycles}
\ee
as it follows from (\ref{gauge1}). Now the key problem of gauge theories is to
prove that
the causal splitting of this commutator can be done such that the gauge
invariance 
property is preserved i.e.
$
A^{IJ}(x,y), R^{IJ}(x,y), T^{IJ}(x,y)
$
are also cocycles. In lower orders of perturbation theory this can be done
elementary.

We address in this paper another question, namely if these objects are in some
sense
coboundaries. We will prove that this is true only for the loop contributions
and 
in the Yang-Mills case only up to super-renormalizable terms.
\newpage

\subsection{Second Order Causal Distributions\label{causal}}

We remind the fact that the Pauli-Villars distribution is defined by
\be
D_{m}(x) = D_{m}^{(+)}(x) + D_{m}^{(-)}(x)
\ee
where 
\be
D_{m}^{(\pm)}(x) \sim 
\int dp e^{i p\cdot x} \theta(\pm p_{0}) \delta(p^{2} - m^{2})
\ee
such that
\be
D^{(-)}(x) = - D^{(+)}(- x).
\ee

This distribution has causal support. In fact, it can be causally split
(uniquely) into an
advanced and a retarded part:
\be
D = D^{\rm adv} - D^{\rm ret}
\ee
and then we can define the Feynman propagator and antipropagator
\be
D^{F} = D^{\rm ret} + D^{(+)}, \qquad \bar{D}^{F} = D^{(+)} - D^{\rm adv}.
\ee
All these distributions have singularity order
$
\omega(D) = -2
$.

For  one-loop contributions in the second order we need the basic distribution
\be
d_{2}(x) \equiv {1\over 2} [ D_{m}^{(+)}(x)^{2} - D_{m}^{(+)}(- x)^{2} ]
\ee
which also with causal support and it can be causally split as above in
\be
d_{2} = d_{2}^{\rm adv} - d_{2}^{\rm ret}
\ee
and the corresponding Feynman propagators can be defined. These distributions
have the singularity order
$
\omega(D) = 0
$.

We will now consider for simplicity the case
$
m = 0.
$

In the explicit computations some associated distributions with causal
support do appear. In the Yang-Mills case we can have two derivatives
distributed in two ways on the two factors
$
D_{0}^{(+)}
$:
\bea
d_{\mu\nu}(x) = D_{0}^{(+)}(x) \partial_{\mu}\partial_{\nu}D_{0}^{(+)}(x) 
- D_{0}^{(+)}(- x) \partial_{\mu}\partial_{\nu}D_{0}^{(+)}(-x)
\nonumber \\
f_{\mu\nu}(x) = \partial_{\mu}D_{0}^{(+)}(x) \partial_{\nu}D_{0}^{(+)}(x) 
- \partial_{\mu}D_{0}^{(+)}(- x) \partial_{\nu}D_{0}^{(+)}(-x).
\eea
It is not hard to prove that we have
\bea
d_{\mu\nu} = 
{2\over 3} \left( \partial_{\mu}\partial_{\nu} - {1\over 4} \square\right)d_{2}
\nonumber \\
f_{\mu\nu} = 
{1\over 3} \left( \partial_{\mu}\partial_{\nu} + {1\over 2} \square\right)d_{2}.
\eea

In the case of gravity we have $4$ derivatives distributed on the two factors
$
D_{0}^{(+)}
$.
This can be done in three different ways and we obtain after some computations
expressions of the type
$
P(\partial)d_{2}
$
where $P$ are polynomials of degree $4$ in the derivatives. As we will see, the
explicit expressions are not needed.
For two-loop contributions in the second order we need 
\be
d_{3}(x) \equiv {1\over 6} [ D_{m}^{(+)}(x)^{3} - D_{m}^{(+)}(- x)^{3} ]
\ee
and the associated distributions
\bea
d_{3}^{(1)}(x) = \partial_{\mu}D_{0}^{(+)}(x) \partial_{\nu}D_{0}^{(+)}(x)
\partial^{\mu}\partial^{\nu}D_{0}^{(+)}(x) 
- \partial_{\mu}D_{0}^{(+)}(- x) \partial_{\nu}D_{0}^{(+)}(-x)
\partial^{\mu}\partial^{\nu}D_{0}^{(+)}(- x) 
\nonumber \\
d_{3}^{(2)}(x) = D_{0}^{(+)}(x) \partial_{\mu}\partial_{\nu}D_{0}^{(+)}(x) 
\partial^{\mu}\partial^{\nu}D_{0}^{(+)}(x) 
- D_{0}^{(+)}(- x) \partial_{\mu}\partial_{\nu}D_{0}^{(+)}(- x) 
\partial^{\mu}\partial^{\nu}D_{0}^{(+)}(-x).
\eea
It can be proved that we have
\be
d_{3}^{(1)}(x) = {1\over 4}~\square^{2} d_{3}(x),
\qquad
d_{3}^{(2)}(x) = {1\over 2}~\square^{2} d_{3}(x).
\ee
\newpage
\section{Yang-Mills Case\label{ym}}
In this section we prove that the one-loop contributions appearing in the
Yang-Mills case are in fact super-renormalizable i.e. the cochains are in fact 
coboundaries, up to super-renormalizable terms. We first need the explicit
expressions for the cochains.
\subsection{The Yang-Mills Lagrangian}

Now we consider the framework and notations from Subsection \ref{generic}. 
Then we have the following result which describes the most general form
of the Yang-Mills interaction \cite{YM}, \cite{standard}, \cite{fermi},
\cite{Sc2}. 
Summation over the dummy indexes is used everywhere.
\begin{thm}
Let $T$ be a relative cocycle which is tri-linear in the fields and is of
canonical dimension
$
\omega(T) \leq 4
$
and null Fermi parity. Then:
(i) $T$ is (relatively) cohomologous to a non-trivial co-cycle of the form:
\bea
T = f_{abc} \left( {1\over 2}~v_{a\mu}~v_{b\nu}~F_{c}^{\nu\mu}
+ u_{a}~v_{b}^{\mu}~\partial_{\mu}\tilde{u}_{c}\right)
\nonumber \\
+ f^{\prime}_{abc} (\Phi_{a}~\phi_{b}^{\mu}~v_{c\mu} +
m_{b}~\Phi_{a}~\tilde{u}_{b}~u_{c})
\nonumber \\
+ {1\over 3!}~f^{\prime\prime}_{abc}~\Phi_{a}~\Phi_{b}~\Phi_{c}
+ j^{\mu}_{a}~v_{a\mu} + j_{a}~\Phi_{a};
\eea
where we can take the constants
$
f_{abc} = 0
$
if one of the indexes is in
$
I_{3};
$
also
$
f^{\prime}_{abc} = 0
$
if 
$
c \in I_{3}
$
or one of the indexes $a$ and $b$ are from
$
I_{1};
$
and
$
j^{\mu}_{a} = 0
$
if
$
a \in I_{3};
$
$
j_{a} = 0
$
if
$
a \in I_{1}.
$
By definition
\be
\phi_{a}^{\mu} \equiv \partial^{\mu}\Phi_{a} - v_{a}^{\mu}
\ee
Moreover we have:

(a) The constants
$
f_{abc}
$
are completely antisymmetric
\be
f_{abc} = f_{[abc]}.
\label{anti-f}
\ee

(b) The expressions
$
f^{\prime}_{abc}
$
are antisymmetric  in the indexes $a$ and $b$:
\be
f^{\prime}_{abc} = - f^{\prime}_{bac}
\label{anti-f'}
\ee
and are connected to 
$f_{abc}$
by:
\be
f_{abc}~m_{c} = f^{\prime}_{cab} m_{a} - f^{\prime}_{cba} m_{b}.
\label{f-f'}
\ee

(c) The (completely symmetric) expressions 
$f^{\prime\prime}_{abc} = f^{\prime\prime}_{\{abc\}}$
verify
\be
f^{\prime\prime}_{abc}~m_{c} = \left\{\begin{array}{rcl} 
{1 \over m_{c}}~f'_{abc}~(m_{a}^{2} - m_{b}^{2}) & \mbox{for} & a, b \in I_{3},
c \in I_{2} \\
- {1 \over m_{c}}~f'_{abc}~m_{b}^{2} & \mbox{for} & a, c \in I_{2}, b \in
I_{3}.\end{array}\right.
\label{f"}
\ee

(d) the expressions
$
j^{\mu}_{a}
$
and
$
j_{a}
$
are bilinear in the Fermi matter fields: in tensor notations;
\bea
j_{a}^{\mu} = \sum_{\epsilon}~
\overline{\psi} t^{\epsilon}_{a} \otimes \gamma^{\mu}\gamma_{\epsilon} \psi
\qquad
j_{a} = \sum_{\epsilon}~
\overline{\psi} s^{\epsilon}_{a} \otimes \gamma_{\epsilon} \psi
\label{current}
\eea
where  for every
$
\epsilon = \pm
$
we have defined the chiral projectors of the algebra of Dirac matrices
$
\gamma_{\epsilon} \equiv {1\over 2}~(I + \epsilon~\gamma_{5})
$
and
$
t^{\epsilon}_{a},~s^{\epsilon}_{a}
$
are 
$
|I_{4}| \times |I_{4}|
$
matrices. If $M$ is the mass matrix
$
M_{AB} = \delta_{AB}~M_{A}
$
then we must have
\be
\partial_{\mu}j^{\mu}_{a} = m_{a}~j_{a} 
\qquad \Leftrightarrow \qquad
m_{a}~s_{a}^{\epsilon} = i(M~t^{\epsilon}_{a} - t^{-\epsilon}_{a}~M).
\label{conserved-current}
\ee

(ii) The relation 
$
d_{Q}T = i~\partial_{\mu}T^{\mu}
$
is verified by:
\be
T^{\mu} = f_{abc} \left( u_{a}~v_{b\nu}~F^{\nu\mu}_{c} -
{1\over 2} u_{a}~u_{b}~d^{\mu}\tilde{u}_{c} \right)
+ f^{\prime}_{abc}~\Phi_{a}~\phi_{b}^{\mu}~u_{c}
+ j^{\mu}_{a}~u_{a}
\label{Tmu}
\ee

(iii) The relation 
$
d_{Q}T^{\mu} = i~\partial_{\nu}T^{\mu\nu}
$
is verified by:
\be
T^{\mu\nu} \equiv {1\over 2} f_{abc}~u_{a}~u_{b}~F_{c}^{\mu\nu}.
\ee

\label{T1}
\end{thm}
\subsection{The Generic Expressions for the One-Loop Cochains}
We consider the one-loop contributions 
$
D_{(1)}^{IJ}(x,y)
$
from
$
D^{IJ}(x,y)
$
and we write for every mass $m$ in the game
\be
D_{m} = D_{0} + ( D_{m} - D_{0})
\label{split}
\ee
In this way we split 
$
D_{(1)}^{IJ}(x,y)
$
into a contribution 
$
D_{(1)0}^{IJ}(x,y)
$
where everywhere
$
D_{m} \mapsto D_{0}
$
and a contribution where at least one factor
$
D_{m}
$
is replaced by the difference
$
D_{m} - D_{0}
$.
Because we have
\be
\omega(D_{m} - D_{0}) = -4
\ee
the second contribution will be super-renormalizable. We now consider the first
contribution.
By direct computations we obtain
\be
D_{(1)0}^{[\mu\nu]\emptyset}(x,y) = 0
\ee
\be
D_{(1)0}^{[\mu][\nu]}(x,y) = (\partial^{\mu}\partial^{\nu} - \eta^{\mu\nu}
\square)d_{2}(x - y) 
\tilde{g}_{ab} u_{a}(x) u_{b}(y)
\ee
\bea
D_{(1)0}^{[\mu] \emptyset}(x,y) = (\partial^{\mu}\partial^{\nu} -
\eta^{\mu\nu}
\square) d_{2}(x - y) 
\tilde{g}_{ab} u_{a}(x) v_{b\nu}(y)
\nonumber \\
+ \partial_{\nu} d_{2}(x - y) g_{ab} [ F^{\mu\nu}_{a}(x) u_{b}(y) - u_{a}(x)
F^{\mu\nu}_{b}(y) ]
\eea
\be
D_{(1)0}^{\emptyset [\mu]}(x,y) = - D_{(1)0}^{[\mu]\emptyset}(y,x)
\ee
\bea
D_{(1)0}^{\emptyset \emptyset}(x,y) = (\partial^{\mu}\partial^{\nu} -
\eta^{\mu\nu}
\square)d_{2}(x - y) 
\tilde{g}_{ab} v_{a\mu}(x) v_{b\nu}(y)
\nonumber \\
+ \partial_{\mu} d_{2}(x - y)
g_{ab} [ - F^{\mu\nu}_{a}(x) v_{b\nu}(y) +  \partial^{\mu}\tilde{u}_{a}(x)
u_{b}(y)
 + v_{a\nu}(x) F^{\mu\nu}_{b}(y) - u_{a}(x)  \partial^{\mu}\tilde{u}_{b}(y) ]
\nonumber \\
- d_{2}(x - y)g_{ab} F^{\mu\nu}_{a}(x) F_{b\mu\nu}(y) 
\nonumber \\
+ \partial_{\mu} d_{2}(x - y)
g^{(3)}_{ab} [ \Phi_{a}(x) \partial^{\mu}\Phi_{b}(y) -
\partial^{\mu}\Phi_{a}(x) \Phi_{b}(y) ]
- 2 d_{2}(x - y) g^{(3)}_{ab} \partial^{\mu}\Phi_{a}(x)
\partial_{\mu}\Phi_{b}(y)
\nonumber \\
- i \partial_{\mu} d_{2}(x - y)
[ \bar{\Psi}(x) A_{\epsilon} \otimes \gamma^{\mu}\gamma_{\epsilon}\Psi(y) 
- \bar{\Psi}(y) A_{\epsilon} \otimes \gamma^{\mu}\gamma_{\epsilon}\Psi(x)]
\nonumber \\
 + \square  d_{2}(x - y) g^{(4)}_{ab} \Phi_{a}(x) \Phi_{b}(y)
\eea
where we have defined some bilinear combinations in the constants appearing in
the 
interaction Lagrangian:
\bea
g_{ab} = f_{pqa} f_{pqb}
\qquad
g^{(1)}_{ab} = f^{\prime}_{pqa} f^{\prime}_{pqb} 
\qquad
g^{(2)}_{ab} = \sum_{\epsilon} Tr(t^{\epsilon}_{a} t^{\epsilon}_{b})
\qquad
g^{(3)}_{ab} = f^{\prime}_{apq} f^{\prime}_{bpq} 
\nonumber \\
g^{(4)}_{ab} = 2 \sum_{\epsilon} Tr(s^{\epsilon}_{a} s^{- \epsilon}_{b})
\qquad
\tilde{g}_{ab} \equiv g_{ab} + {1\over 2} g^{(1)}_{ab} + 2 g^{(2)}_{ab}
\qquad
A_{\epsilon} = \sum_{a} ( 2 t^{\epsilon}_{a} t^{\epsilon}_{a} +
s^{-\epsilon}_{a} s^{\epsilon}_{a}).
\eea
\subsection{The Generic Form of the Coboundaries}
In this Subsection we prove the basic result for the Yang-Mills case.
\begin{thm}
The expression
$
D_{(1)0}^{IJ}(x,y)
$
is a coboundary
\be
D_{(1)0}^{IJ} = (\bar{s}B)^{IJ}.
\ee
\end{thm}
{\bf Proof:}
It is done by providing an explicit expression of the coboundary. If we define
\be
B^{[\mu\nu][\rho]}(x,y) = (\eta^{\mu\rho}\partial^{\nu} - \eta^{\nu\rho}
\partial^{\mu})d_{2}(x - y) 
h^{(0)}_{ab} u_{a}(x) u_{b}(y)
\ee
\bea
B^{[\mu\nu]\emptyset}(x,y) = 
(\eta^{\mu\rho}\partial^{\nu} - \eta^{\nu\rho} \partial^{\mu})d_{2}(x - y) 
h^{(0)}_{ab} u_{a}(x) v_{b\rho}(y)
\nonumber \\
+ d_{2}(x - y) h^{(1)}_{ab} F^{\mu\nu}_{a}(x) u_{b}(y) 
+ d_{2}(x - y) h^{(2)}_{ab}  u_{a}(x) F^{\mu\nu}_{b}(y) 
\eea
\bea
B^{[\mu][\nu]}(x,y) = (\eta^{\nu\rho}\partial^{\mu} - \eta^{\mu\nu}
\partial^{\rho}) d_{2}(x - y) 
h^{(0)}_{ab} v_{a\rho}(x) u_{b}(y)
\nonumber \\
+ (\eta^{\mu\rho}\partial^{\nu} - \eta^{\mu\nu} \partial^{\rho}) d_{2}(x - y) 
h^{(0)}_{ab} u_{a}(x) v_{b\rho}(y)
\nonumber \\
+ d_{2}(x - y) h^{(3)}_{ab} [ F^{\mu\nu}_{a}(x) u_{b}(y) + u_{a}(x)
F^{\mu\nu}_{b}(y) 
\eea
\bea
B^{[\mu] \emptyset}(x,y) = 
(\eta^{\nu\rho}\partial^{\mu} - \eta^{\mu\rho} \partial^{\nu}) d_{2}(x - y) 
h^{(0)}_{ab} v_{a\nu}(x) v_{b\rho}(y)
\nonumber \\
+ d_{2}(x - y) h^{(1)}_{ab} \partial^{\mu}\tilde{u}_{a}(x) u_{b}(y) 
- d_{2}(x - y) h^{(2)}_{ab}  v_{a\nu}(x) F^{\mu\nu}_{b}(y)
\nonumber\\
+ d_{2}(x - y) h^{(3)}_{ab} 
[ F^{\mu\nu}_{a}(x) v_{b\nu}(y) - u_{a}(x) \partial^{\mu}\tilde{u}_{b}(y) ]
\nonumber \\
+ d_{2}(x - y) h^{(4)}_{ab} \Phi_{a}(x) \partial^{\mu}\Phi_{b}(y) 
+ d_{2}(x - y) h^{(5)}_{ab}  \partial^{\mu}\Phi_{a}(x) \Phi_{b}(y)
\nonumber \\
- {i \over 2}~g^{(4)}_{ab}~[ \partial^{\mu}d_{2}(x - y) \Phi_{a}(x) \Phi_{b}(y)
- d_{2}(x - y) \partial^{\mu}\Phi_{a}(x) \Phi_{b}(y) ]
\nonumber \\
- d_{2}(x - y) \bar{\Psi}(x) A_{\epsilon} \otimes
\gamma^{\mu}\gamma_{\epsilon}\Psi(y)
\eea
\be
B^{\emptyset\emptyset}(x, y) = - d_{2}(x - y) h^{(3)}_{ab}
[ \partial_{\mu}\tilde{u}_{a}(x) v_{b}^{\mu}(y) - (x \leftrightarrow y) ]
\ee
where
\bea
h^{(0)} = - {i \over 2} \tilde{g}
\qquad
h^{(1)} = - i \left( {1\over 2} \tilde{g} + 3 g\right)
\nonumber \\
h^{(2)} = - i g
\qquad
h^{(3)} = i \left( {1\over 2} \tilde{g} + 2 g\right)
\nonumber \\
h^{(4)} = i g^{(3)}
\qquad
h^{(5)} = 2 i g^{(3)}
\eea
then we can prove the relation from the statement. As a result we have:
\begin{thm}
The one-loop contribution
$
D_{(1)}^{IJ}(x,y)
$
is a coboundary, up to super-renormalizable terms. So the causal commutator 
$
D^{IJ}(x,y)
$
up to the second-order of the perturbation theory is the sum of the classical
contribution
(tree part) and quantum (loop) corrections which are super-renormalizable. This 
property remains true after causal decomposition, in particular for the
chronological
products.
\end{thm}
For the two-loop contribution in the second order of the perturbation theory we
find 
the non-trivial expression
\be
D^{\emptyset\emptyset}_{(2)0}(x,y) = i c~\square d_{3}(x - y)
\ee
(where $c$ is a constant). If we take 
\bea
B^{\emptyset\emptyset}_{(2)}(x,y) = 0
\nonumber\\
B^{[\mu]\emptyset}_{(2)}(x,y) = {1\over 2} c~\partial^{\mu}d_{3}(x - y)
\eea
then we can express the commutator
$
D^{\emptyset\emptyset}_{(2)0}(x,y)
$
as a coboundary.
\newpage
\section{Gravity\label{gravity}}
We prove a result of the same nature for the extremely interesting case of
massless gravity.
\subsection{The Gravity Lagrangian}
We have the following result \cite{cohomology2}.
\begin{thm}
Let $T$ be a relative cocycle for 
$
d_{Q}
$
which is tri-linear in the fields and is of canonical dimension
$
\omega(T) \leq 5
$
and ghost number
$
gh(T) = 0.
$
Then:
(i) $T$ is (relatively) cohomologous to a non-trivial cocycle of the form:
\bea
T = \kappa (
2~h_{\mu\rho}~\partial^{\mu}h^{\nu\lambda}~\partial^{\rho}h_{\nu\lambda}
+ 4~h_{\nu\rho}~\partial^{\lambda}h^{\mu\nu}~\partial_{\mu}{h_{\nu}}^{\lambda}
- 4~h_{\rho\lambda}~\partial^{\mu}h^{\nu\rho}~\partial_{\mu}{h_{\nu}}^{\lambda}
\nonumber \\
+ 2~h^{\rho\lambda}~\partial_{\mu}h_{\rho\lambda}~\partial^{\mu}h
- h_{\mu\rho}~\partial^{\mu}h~d^{\rho}h 
- 4~u^{\rho}~\partial^{\nu}\tilde{u}^{\lambda}~\partial_{\rho}h_{\nu\lambda}
\nonumber \\
+ 4~\partial^{\rho}u^{\nu}~\partial_{\nu}\tilde{u}^{\lambda}~h_{\rho\lambda}
+ 4~\partial^{\rho}u_{\nu}~\partial^{\lambda}\tilde{u}_{\nu}~h_{\rho\lambda}
- 4~\partial^{\nu}u_{\nu}~\partial^{\rho}\tilde{u}^{\lambda}~h_{\rho\lambda})
\eea
where
$
\kappa \in \R.
$

(ii) The relation 
$
d_{Q}T = i~\partial_{\mu}T^{\mu}
$
is verified by:
\bea
T^{\mu} = \kappa ( - 2
u^{\mu}~\partial_{\nu}h_{\rho\lambda}~\partial^{\rho}h^{\nu\lambda} 
+ u^{\mu}~\partial_{\rho}h_{\nu\lambda}~\partial^{\rho}h^{\nu\lambda} 
- {1\over 2} u^{\mu}~\partial_{\rho}h~\partial^{\rho}h
\nonumber \\
+ 4~u^{\rho}~\partial^{\nu}h^{\mu\lambda}~\partial_{\rho}h_{\nu\lambda}
- 2~u^{\rho}~\partial^{\mu}h^{\nu\lambda}~\partial_{\rho}h_{\nu\lambda}
+ u^{\rho}~\partial^{\mu}h~\partial_{\rho}h
\nonumber \\
- 4~\partial^{\rho}u^{\nu}~\partial_{\nu}h^{\mu\lambda}~h_{\rho\lambda}
- 4\partial~^{\rho}u_{\nu}~\partial^{\lambda}h^{\mu\nu}~h_{\rho\lambda}
+ 4~\partial^{\lambda}u_{\rho}~\partial^{\mu}h^{\nu\rho}~h_{\nu\lambda}
\nonumber \\
+ 4~\partial_{\nu}u^{\nu}~\partial^{\rho}h^{\mu\lambda}~h_{\rho\lambda}
- 2~\partial_{\nu}u^{\nu}~\partial^{\mu}h^{\rho\lambda}~h_{\rho\lambda}
- 2~\partial^{\rho}u^{\lambda}~h_{\rho\lambda}~\partial^{\mu}h
+ \partial^{\nu}u_{\nu}~h~\partial^{\mu}h
\nonumber \\
- 2~u^{\mu}~\partial_{\nu}\partial_{\rho}u^{\rho}~\tilde{u}^{\nu}
+ 2~u_{\rho}~\partial^{\rho}\partial^{\sigma}u_{\sigma}~\tilde{u}^{\mu}
- 2~u^{\mu}~\partial_{\lambda}u_{\rho}~\partial^{\rho}\tilde{u}^{\nu}
\nonumber \\
+ 2~u_{\rho}~\partial_{\lambda}u^{\mu}~\partial^{\rho}\tilde{u}^{\lambda}
+ 2~\partial^{\rho}u_{\rho}~\partial_{\lambda}u^{\mu}~\tilde{u}^{\lambda}
- 2~u_{\rho}~\partial^{\rho}u_{\lambda}~\partial^{\mu}\tilde{u}^{\lambda})
\label{Tmu-gr}
\eea

(iii) The relation 
$
d_{Q}T^{\mu} = i~\partial_{\nu}T^{\mu\nu}
$
is verified by:
\bea
T^{\mu\nu} \equiv \kappa [ 2 ( -
u^{\mu}~\partial_{\lambda}u_{\rho}~\partial^{\rho}h^{\nu\lambda}
+ u_{\rho}~\partial_{\lambda}u^{\mu}~\partial^{\rho}h^{\nu\lambda}
+ u_{\rho}~\partial^{\rho}u_{\lambda}~\partial^{\nu}h^{\mu\lambda}
+ \partial_{\rho}u^{\rho}~\partial_{\lambda}u^{\mu}~h^{\nu\lambda})
\nonumber \\
- (\mu \leftrightarrow \nu)
+ 4~\partial^{\lambda}u^{\mu}~\partial^{\rho}u^{\nu}~h_{\rho\lambda} ].
\label{Tmunu}
\eea

(iv) The relation 
$
d_{Q}T^{\mu\nu} = i~\partial_{\rho}T^{\mu\nu\rho}
$
is verified by:
\bea
T^{\mu\nu\rho} \equiv \kappa [ 2
u_{\lambda}~\partial^{\lambda}u^{\rho}~u^{\mu\nu}
- u_{\rho}~(\partial^{\mu}u^{\lambda}~\partial_{\lambda}u^{\nu}
- \partial^{\nu}u^{\lambda}~\partial_{\lambda}u^{\mu}) 
+ {\rm circular~perm.}]
\label{Tmunurho}
\eea
and we have
$
d_{Q}T^{\mu\nu\rho} = 0.
$

(v) The cocycles
$
T, T^{\mu}, T^{\mu\nu}
$
and
$
T^{\mu\nu\rho}
$
are non-trivial and invariant with respect to parity. Here
\be
u_{\mu\nu} \equiv \partial_{\mu}u_{\nu} - \partial_{\nu}u_{\mu}.
\ee
\label{T1gravity}
\end{thm}
\subsection{The Generic Expressions for the One-Loop Cochains}
We consider the one-loop contribution
$
D_{(1)}^{IJ}(x,y)
$
and we do not need the splitting (\ref{split}) of the Pauli-Jordan causal
commutator from the preceding Section because the mass is already null. There is
a particularity of the gravity case, namely we do not need to compute explicitly
these expressions. The result is of pure cohomology nature. We only need to
provide a generic expression for 
$
D_{(1)}^{IJ}(x,y)
$
and impose the cochain condition
\be
sD = 0
\ee
which follows from the gauge invariance of the interaction Lagrangian
$
sT = 0.
$
A number of limitations will result on the various arbitrary coefficients and we
will be able to prove that $D$ is a coboundary. We give only the relevant
coefficients.
For instance
from the expression
$
D^{[\mu]\emptyset}
$
we need only:
\bea
D^{[\mu]\emptyset}(x,y) = \cdots
+ F_{3}~\partial^{\mu}\partial^{\nu}\square d_{2}(x-y)~
u^{\rho}(x)~h_{\nu\rho}(y) + \cdots
\nonumber \\
+ F_{12}~\partial_{\rho}\partial_{\sigma}\square d_{2}(x-y)~
h^{\mu\sigma}(x)~u^{\rho}(y)
\nonumber \\
+ F_{13}~\square^{2}d_{2}(x-y)~u_{\nu}(x)~h^{\mu\nu}(y)
+ F_{14}~\square^{2}d_{2}(x-y)~h^{\mu\nu}(x)~u_{\nu}(y)
\eea
From the expression
$
D^{[\mu][\nu]}
$
we need the whole sector 
\bea
D^{[\mu][\nu]}_{1}(x,y) =
K_{1}~\partial^{\mu}\partial^{\nu}\partial^{\rho}\partial^{\sigma}
d_{2}(x-y)~u_{\rho}(x)~u_{\sigma}(y)
\nonumber \\
~+ K_{2}~\partial^{\mu}\partial^{\nu}\square d_{2}(x-y)~u_{\rho}(x)~u^{\rho}(y)
\nonumber \\
~+ K_{3}~[
\partial^{\mu}\partial^{\rho}\square d_{2}(x-y)~u_{\rho}(x)~u^{\nu}(y) +
\partial^{\nu}\partial^{\rho}~\square d_{2}(x-y)~u^{\mu}(x)~u_{\rho}(y) ]
\nonumber \\
~+ K_{4}~[
\partial^{\mu}\partial^{\rho}\square d_{2}(x-y)~u^{\nu}(x)~u_{\rho}(y) +
\partial^{\nu}\partial^{\rho}~\square d_{2}(x-y)~u_{\rho}(x)~u^{\mu}(y) ]
\nonumber \\
~+ K_{5}~\eta^{\mu\nu}~
\partial^{\rho}\partial^{\sigma}\square d_{2}(x-y)~u_{\rho}(x)~u_{\sigma}(y)
\nonumber \\
~+ K_{6}~\eta^{\mu\nu}~\square^{2} d_{2}(x-y)~u_{\rho}(x)~u^{\rho}(y)
\nonumber \\
+ K_{7}~\square^{2} d_{2}(x-y)~u^{\mu}(x)~u^{\nu}(y)
\nonumber \\
+ K_{8}~\square^{2} d_{2}(x-y)~u^{\nu}(x)~u^{\mu}(y)
\eea
and only a few terms from the sector:
\bea
D^{[\mu][\nu]}_{2}(x,y) = \cdots
+ L_{2}~\partial^{\mu}\partial^{\nu}\partial^{\rho}d_{2}(x-y)~
[u^{\sigma}(x)~\partial_{\rho}u_{\sigma}(y) -
\partial_{\rho}u_{\sigma}(x)~u^{\sigma}(y) ]
\nonumber \\
+ L_{3}~\partial^{\mu}\partial^{\nu}\partial^{\rho}d_{2}(x-y)~
[u^{\sigma}(x)~\partial_{\sigma}u_{\rho}(y) -
\partial_{\sigma}u_{\rho}(x)~u^{\sigma}(y) ] + \cdots
\nonumber \\
+ L_{5}~[\partial^{\mu}\partial^{\rho}\partial^{\sigma}d_{2}(x-y)~
u_{\rho}(x)~\partial^{\nu}u_{\sigma}(y) 
- \partial^{\nu}\partial^{\rho}\partial^{\sigma}d_{2}(x-y)~
\partial^{\mu}u_{\sigma}(x)~u_{\rho}(y) ]
\nonumber \\
+ L_{6}~[\partial^{\mu}\partial^{\rho}\partial^{\sigma}d_{2}(x-y)~
u_{\rho}(x)~\partial_{\sigma}u^{\nu}(y) 
- \partial^{\nu}\partial^{\rho}\partial^{\sigma}d_{2}(x-y)~
\partial_{\sigma}u^{\mu}(x)~u_{\rho}(y) ] + \cdots
\nonumber \\
+ L_{8}~[
\partial^{\mu}\square d_{2}(x-y)~u_{\rho}(x)~\partial^{\nu}u^{\rho}(y) -
\partial^{\nu}~\square d_{2}(x-y)~\partial^{\mu}u^{\rho}(x)~u_{\rho}(y) ]
\nonumber \\
~+ L_{9}~[
\partial^{\mu}\square d_{2}(x-y)~u_{\rho}(x)~\partial^{\rho}u^{\nu}(y) -
\partial^{\nu}~\square d_{2}(x-y)~\partial^{\rho}u^{\mu}(x)~u_{\rho}(y) ] +
\nonumber \\
+ L_{10}~\partial_{\rho}\square d_{2}(x-y)~
[u^{\rho}(x)~\partial^{\mu}u^{\nu}(y) - \partial^{\nu}u^{\mu}(x)~u_{\rho}(y) ]
\nonumber \\
+ L_{11}~\partial_{\rho}\square d_{2}(x-y)~
[u^{\rho}(x)~\partial^{\nu}u^{\mu}(y) - \partial^{\mu}u^{\nu}(x)~u_{\rho}(y) ]
\nonumber \\
+ L_{12}~\partial_{\rho}\square d_{2}(x-y)~
[u^{\mu}(x)~\partial^{\rho}u^{\nu}(y) - \partial^{\rho}u^{\mu}(x)~u_{\nu}(y) ]
\nonumber \\
+ L_{13}~\partial_{\rho}\square d_{2}(x-y)~
[u^{\mu}(x)~\partial^{\nu}u^{\rho}(y) - \partial^{\mu}u^{\rho}(x)~u_{\nu}(y) ]
+ \cdots
\nonumber \\
+ L_{18}~\eta^{\mu\nu}~\partial^{\rho}\square d_{2}(x-y)~
[u^{\sigma}(x)~\partial_{\rho}u_{\sigma}(y) -
\partial_{\rho}u_{\sigma}(x)~u^{\sigma}(y) ]
\nonumber\\
+ L_{19}~\eta^{\mu\nu}~\partial^{\rho}\square d_{2}(x-y)~
[u^{\sigma}(x)~\partial_{\sigma}u_{\rho}(y) -
\partial_{\sigma}u_{\rho}(x)~u^{\sigma}(y) ] + \cdots
\nonumber\\
~+ L_{24}~[
\partial^{\nu}\square d_{2}(x-y)~u_{\rho}(x)~\partial^{\mu}u^{\rho}(y) -
\partial^{\mu}~\square d_{2}(x-y)~\partial^{\nu}u^{\rho}(x)~u_{\rho}(y) ]
\nonumber \\
~+ L_{25}~[
\partial^{\nu}\square d_{2}(x-y)~u_{\rho}(x)~\partial^{\rho}u^{\mu}(y) -
\partial^{\mu}~\square d_{2}(x-y)~\partial^{\rho}u^{\nu}(x)~u_{\rho}(y) ]  +
\cdots
\eea
From the expression
$
D^{[\mu\nu]\emptyset}
$
we need the whole sector
\bea
D^{[\mu\nu]\emptyset}_{1}(x,y) =
Q_{1}~[ \partial^{\mu}\partial^{\rho}\square d_{2}(x-y)~u^{\nu}(x)~u_{\rho}(y)
- (\mu \leftrightarrow \nu) ]
\nonumber \\
+ Q_{2}~[ \partial^{\mu}\partial^{\rho}\square d_{2}(x-y)~u_{\rho}(x)~u^{\nu}(y)
- (\mu \leftrightarrow \nu) ]
\nonumber \\
+ Q_{3}~\square^{2} d_{2}(x-y)~[ u^{\mu}(x)~u^{\nu}(y) 
- (\mu \leftrightarrow \nu) ]
\eea
and some terms from the sector
\bea
D^{[\mu\nu]\emptyset}_{2}(x,y) = \cdots
+ R_{5}~[\partial^{\mu}\square d_{2}(x-y)~u_{\rho}(x)~\partial^{\nu}u^{\rho}(y)
- (\mu \leftrightarrow \nu)  ]
\nonumber \\
~+ R_{6}~[\partial^{\mu}\square d_{2}(x-y)~u_{\rho}(x)~\partial^{\rho}u^{\nu}(y)
- (\mu \leftrightarrow \nu) ]  +
\cdots
\eea
and from
\bea
D^{[\mu\nu]\emptyset}_{3}(x,y) = \cdots
+ S_{2}~[ \partial^{\mu}\partial^{\rho}\partial^{\sigma}d_{2}(x-y)~
\partial^{\nu}u_{\rho}(x)~u_{\sigma}(y) - (\mu \leftrightarrow \nu)  ]
\nonumber \\
+ S_{3}~[ \partial^{\mu}\partial^{\rho}\partial^{\sigma}d_{2}(x-y)~
\partial_{\rho}u^{\nu}(x)~u_{\sigma}(y) - (\mu \leftrightarrow \nu)  ]
+ \cdots
\nonumber \\
+ S_{5}~[\partial^{\mu}\square d_{2}(x-y)~\partial^{\nu}u^{\rho}(x)~u_{\rho}(y)
- (\mu \leftrightarrow \nu)  ]
\nonumber \\
+ S_{6}~[\partial^{\mu}\square d_{2}(x-y)~\partial^{\rho}u^{\nu}(x)~u_{\rho}(y)
- (\mu \leftrightarrow \nu) 
\nonumber \\
+ S_{7}~\partial_{\rho}\square d_{2}(x-y)~
[\partial^{\mu}u^{\nu}(x) - (\mu \leftrightarrow \nu) ]~u^{\rho}(y)
\nonumber \\
+ S_{8}~\partial_{\rho}\square d_{2}(x-y)~
\partial^{\rho}u^{\mu}(x)~u^{\nu}(y) - (\mu \leftrightarrow \nu)
\nonumber \\
+ S_{9}~\partial_{\rho}\square d_{2}(x-y)~
\partial^{\mu}u^{\rho}(x)~u^{\nu}(y) - (\mu \leftrightarrow \nu)
\eea

\subsection{Relative Cocycle Equations}
Now we consider the cocycle equation
\be
sD = 0 \qquad \Leftrightarrow \qquad
d_{Q}D^{IJ} = i {\partial\over \partial x^{\rho}}D^{I\rho,J} 
+ i (-1)^{|I|} {\partial\over \partial y^{\rho}}D^{I,J\rho}.
\ee
From
\be
d_{Q}D^{[\mu]\emptyset} = 
i {\partial\over \partial x^{\nu}}D^{[\mu\nu]\emptyset} 
- i {\partial\over \partial y^{\nu}}D^{[\mu][\nu]} 
\ee
we obtain:

- from the coefficients of the monomials
$
\partial\partial\partial\partial\partial d_{2}(x - y) u(x) u(y)
$
\be
K_{1} + K_{3} + K_{4} + K_{5} + Q_{1} + Q_{2} = 0
\label{(11.1)}
\ee
\be
K_{2} + K_{6} = 0
\label{(11.2)}
\ee
\be
K_{3} + K_{7} - Q_{1} + Q_{3} = 0
\label{(11.3)}
\ee
\be
K_{4} + K_{8} - Q_{2} - Q_{3} = 0
\label{(11.4)}
\ee  

From (\ref{(11.1)}) + (\ref{(11.3)}) + (\ref{(11.4)}) we obtain
\be
K_{1} +  2 K_{3} + 2 K_{4} + K_{5} + K_{7} + K_{8} = 0
\label{(11)}
\ee 

- from the coefficients of the monomials
$
\partial\partial\partial\partial d_{2}(x - y) u(x) \partial u(y)
$
\be
- K_{2} + L_{2} + L_{8} + L_{18} + R_{5} = {1\over 2}~F_{3}
\label{(12.2)}
\ee
\be
- K_{4} + L_{3} + L_{9} + L_{19} + R_{6} = {1\over 2}~F_{3}
\label{(12.5)}
\ee
\be
- K_{6} + L_{24} - R_{5} = {1\over 2}~F_{13}
\label{(12.8)}
\ee
\be
- K_{8} + L_{25} - R_{6} = {1\over 2}~F_{13}
\label{(12.10)}
\ee

Taking the difference we obtain
\be
- K_{2} + K_{4} + L_{2} - L_{3} + L_{8} - L_{9} + L_{18} - L_{19} 
+ R_{5} - R_{6} = 0
\label{(12.2a)}
\ee
\be
K_{6} - K_{8} - L_{24} + L_{25} + R_{5} - R_{6} = 0
\label{(12.8a)}
\ee
If we subtract these equations and use (\ref{(11.2)}) we get
\be
K_{4} + K_{8} + L_{2} - L_{3} + L_{8} - L_{9} + L_{18} - L_{19} 
+ L_{24} - L_{25} = 0
\label{(12.2b)}
\ee

- from the coefficients of the monomials
$
\partial\partial\partial\partial d_{2}(x - y) \partial u(x) u(y)
$
\be
- L_{5} - L_{11} - L_{13} - S_{2} + S_{7} + S_{9} = - {1\over 2}~F_{12}
\label{(13.5)}
\ee
\be
- L_{6} - L_{10} - L_{12} - Q_{1} - S_{3} - S_{7} + S_{8} = - {1\over 2}~F_{12}
\label{(13.6)}
\ee
\be
- L_{8} - S_{5} = - {1\over 2}~F_{14}
\label{(13.8)}
\ee
\be
- L_{9} + Q_{3} - S_{6} = - {1\over 2}~F_{14}
\label{(13.9)}
\ee

Taking the difference we obtain
\be
- L_{5} + L_{6} + L_{10}- L_{11} + L_{12} - L_{13} +  Q_{1} - S_{2} + S_{3} 
+ 2 S_{7} - S_{8} + S_{9} = 0
\label{(13.5a)}
\ee
\be
L_{8} - L_{9} + Q_{3} + S_{5} - S_{6} = 0
\label{(13.8a)}
\ee
If we add the first equation with (\ref{(11.3)}) and (\ref{(11.4)}) we obtain
\be
K_{3} + K_{4} + K_{7} + K_{8} - L_{5} + L_{6} + L_{10}- L_{11} + L_{12} - L_{13}
- Q_{2} - S_{2} + S_{3} + 2 S_{7} - S_{8} + S_{9} = 0
\label{(13.5b)}
\ee
and if we add the second equation with (\ref{(11.4)}) we obtain
\be
K_{4} + K_{8} + L_{8} - L_{9} -Q_{2}  + S_{5} - S_{6} = 0.
\label{(13.8b)}
\ee

\subsection{The Generic Expressions for the Coboundaries $B^{IJ}$}
\begin{thm}
The coboundary equation
\be
D^{IJ} = (\bar{s}B)^{IJ}, \qquad |I| + |J| = 2
\ee
is true iff the coefficients of the left hand side verify:
\bea
K_{1} +  2 K_{3} + 2 K_{4} + K_{5} + K_{7} + K_{8} = 0
\nonumber \\
K_{2} + K_{6} = 0
\nonumber \\
K_{4} + K_{8} - Q_{2} - Q_{3} = 0
\nonumber \\
K_{3} + K_{7} - Q_{1} + Q_{3} = 0
\nonumber \\
K_{6} - K_{8} - L_{24} + L_{25} + R_{5} - R_{6} = 0
\nonumber \\
K_{4} + K_{8} + L_{8} - L_{9} -Q_{2}  + S_{5} - S_{6} = 0
\nonumber \\
K_{4} + K_{8} + L_{2} - L_{3} + L_{8} - L_{9} + L_{18} - L_{19} 
+ L_{24} - L_{25} = 0
\nonumber \\
K_{3} + K_{4} + K_{7} + K_{8} - L_{5} + L_{6} + L_{10}- L_{11} + L_{12} - L_{13}
\nonumber \\
- Q_{2} - S_{2} + S_{3} + 2 S_{7} - S_{8} + S_{9} = 0.
\eea
\end{thm}
{\bf Proof:}
(i) We need the generic form of the cocycles
$
B^{IJ}
$
constrained by 
\be
gh(B^{IJ}) = |I| + |J| - 1, \qquad \omega(B^{IJ}) = 5.
\ee
We will give only a number of relevant terms:
\bea
B^{[\mu\nu][\rho]}_{1}(x,y) =
a_{1} [\partial^{\mu}\partial^{\rho}\partial^{\sigma} d_{2}(x-y)
~u^{\nu}(x)~u_{\sigma}(y) - (\mu \leftrightarrow \nu) ]
\nonumber \\
~+ a_{2}~[ \partial^{\mu}\partial^{\rho}\partial^{\sigma} d_{2}(x-y)
~u_{\sigma}(x)~u^{\nu }(y) - (\mu \leftrightarrow \nu) ]
\nonumber \\
~+ a_{3}~[\partial^{\mu}\square d_{2}(x-y)~u^{\nu}(x)~u^{\rho}(y) 
- (\mu \leftrightarrow \nu) ]
\nonumber \\
~+ a_{4}~[\partial^{\mu}\square d_{2}(x-y)~u^{\rho}(x)~u^{\nu}(y) 
- (\mu \leftrightarrow \nu)]
\nonumber \\
~+ a_{5}~[ \partial^{\rho}\square d_{2}(x-y)~u^{\mu}(x)~u^{\nu}(y)
- (\mu \leftrightarrow \nu) ]
\nonumber \\
~+ a_{6}~[\eta^{\mu\rho}~\partial^{\nu}\square d_{2}(x-y)
u_{\sigma}(x)~u^{\sigma}(y)- (\mu \leftrightarrow \nu) ]
\nonumber \\
+ a_{7}~[\eta^{\mu\rho}~\partial^{\sigma}\square d_{2}(x-y)
u^{\nu}(x)~u_{\sigma}(y)- (\mu \leftrightarrow \nu) ]
\nonumber \\
+ a_{8}~[\eta^{\mu\rho}~\partial^{\sigma}\square d_{2}(x-y)
u_{\sigma}(x)~u^{\nu}(y)- (\mu \leftrightarrow \nu) ]
\nonumber \\
+ a_{9}~[\eta^{\mu\rho}~
\partial^{\nu}\partial^{\sigma}\partial^{\lambda} d_{2}(x-y)
u_{\sigma}(x)~u_{\lambda}(y)- (\mu \leftrightarrow \nu) ]
\eea

\bea
B^{[\mu\nu][\rho]}_{2}(x,y) =
b_{1} [\partial^{\mu}\partial^{\rho} d_{2}(x-y)
~\partial^{\nu}u^{\sigma}(x)~u_{\sigma}(y) - (\mu \leftrightarrow \nu) ]
\nonumber \\
~+ b_{2} [\partial^{\mu}\partial^{\rho} d_{2}(x-y)
~u_{\sigma}(x)~\partial^{\nu}u^{\sigma}(y) - (\mu \leftrightarrow \nu) ]
\nonumber \\
~+ b_{3}~[ \partial^{\mu}\partial^{\rho} d_{2}(x-y)
~\partial^{\sigma}u^{\nu}(x)~u_{\sigma }(y) - (\mu \leftrightarrow \nu) ]
\nonumber \\
~+ b_{4}~[ \partial^{\mu}\partial^{\rho} d_{2}(x-y)
~u_{\sigma }(x)~\partial^{\sigma}u^{\nu}(y) - (\mu \leftrightarrow \nu) ]
+ \cdots
\nonumber \\
+ b_{7}~[ \partial^{\mu}\partial^{\sigma} d_{2}(x-y)
~\partial^{\nu}u^{\rho}(x)~u_{\sigma}(y) - (\mu \leftrightarrow \nu) ]
+ \cdots
\nonumber \\
+ b_{9}~[ \partial^{\mu}\partial^{\sigma} d_{2}(x-y)
~\partial^{\nu}u_{\sigma}(x)~u^{\rho}(y) - (\mu \leftrightarrow \nu) ]
\nonumber \\
+ b_{11}~[ \partial^{\mu}\partial^{\sigma} d_{2}(x-y)
~\partial^{\rho}u^{\nu}(x)~u_{\sigma}(y) - (\mu \leftrightarrow \nu) ]
\nonumber \\
+ b_{13}~[ \partial^{\mu}\partial^{\sigma} d_{2}(x-y)
~\partial_{\sigma}u^{\nu}(x)~u^{\rho}(y) - (\mu \leftrightarrow \nu) ]
+ \cdots
\nonumber \\
+ b_{19}~\partial^{\rho}\partial^{\sigma} d_{2}(x-y)
~[ \partial^{\mu}u^{\nu}(x)~u_{\sigma}(y) - (\mu \leftrightarrow \nu) ]
+ \cdots
\nonumber \\
+ b_{21}~\partial^{\rho}\partial^{\sigma} d_{2}(x-y)
~[\partial^{\mu}u_{\sigma}(x)~u^{\nu}(y) - (\mu \leftrightarrow \nu) ]
\nonumber \\
+ b_{22}~\partial^{\rho}\partial^{\sigma} d_{2}(x-y)
~[ u^{\mu}(x)~\partial^{\nu}u_{\sigma}(y) - (\mu \leftrightarrow \nu) ]
\nonumber \\
+ b_{23}~\partial^{\rho}\partial^{\sigma} d_{2}(x-y)
~[ \partial_{\sigma}u^{\mu}(x)~u^{\nu}(y) - (\mu \leftrightarrow \nu) ]
+ \cdots
\nonumber \\
+ b_{25}~\square d_{2}(x-y)
~[ \partial^{\mu}u^{\nu}(x)~u^{\rho}(y) - (\mu \leftrightarrow \nu) ]
+ \cdots
\nonumber \\
+ b_{27}~\square d_{2}(x-y)
~[ \partial^{\mu}u^{\rho}(x)~u^{\nu}(y) - (\mu \leftrightarrow \nu) ]
+ \cdots
\nonumber \\
+ b_{29}~\square d_{2}(x-y)
~[ \partial^{\rho}u^{\mu}(x)~u^{\nu}(y) - (\mu \leftrightarrow \nu) ]
+ \cdots
\nonumber \\
+ b_{31}~[ \eta^{\mu\rho}~\partial^{\nu}\partial^{\sigma} d_{2}(x-y)~
\partial_{\sigma}u_{\lambda}(x)~u^{\lambda}(y) - (\mu \leftrightarrow \nu) ]
\nonumber \\
+ b_{32}~[ \eta^{\mu\rho}~\partial^{\nu}\partial^{\sigma} d_{2}(x-y)~
u^{\lambda}(x)~\partial_{\sigma}u_{\lambda}(y) - (\mu \leftrightarrow \nu) ]
\nonumber \\
+ b_{33}~[ \eta^{\mu\rho}~\partial^{\nu}\partial^{\sigma} d_{2}(x-y)~
\partial_{\lambda}u_{\sigma}(x)~u^{\lambda}(y) - (\mu \leftrightarrow \nu) ]
\nonumber \\
+ b_{34}~[ \eta^{\mu\rho}~\partial^{\nu}\partial^{\sigma} d_{2}(x-y)~
u^{\lambda}(x)~\partial_{\lambda}u_{\sigma}(y) - (\mu \leftrightarrow \nu) ]
+ \cdots
\nonumber \\
+ b_{37}~ \partial^{\sigma}\partial^{\lambda} d_{2}(x-y)~[ \eta^{\mu\rho}~
~\partial^{\nu}u_{\sigma}(x)~u_{\lambda}(y) - (\mu \leftrightarrow \nu) ]
+ \cdots
\nonumber \\
+ b_{39}~\partial^{\sigma}\partial^{\lambda} d_{2}(x-y)~[ \eta^{\mu\rho}~
~\partial_{\sigma}u^{\nu}(x)~u_{\lambda}(y) - (\mu \leftrightarrow \nu) ]
+ \cdots
\nonumber \\
~+ b_{43}~\square d_{2}(x-y)~[\eta^{\mu\rho}~
\partial^{\nu}u^{\sigma}(x)~u_{\sigma}(y) - (\mu \leftrightarrow \nu) ]
\nonumber \\
~+ b_{44}~\square d_{2}(x-y)~[\eta^{\mu\rho}~
u_{\sigma}(x)~\partial^{\nu}u^{\sigma}(y) - (\mu \leftrightarrow \nu) ]
\nonumber \\
~+ b_{45}~\square d_{2}(x-y)~[\eta^{\mu\rho}~
\partial^{\sigma}u^{\nu}(x)~u_{\sigma}(y) - (\mu \leftrightarrow \nu) ] 
\nonumber \\
~+ b_{46}~\square d_{2}(x-y)~[\eta^{\mu\rho}~
u_{\sigma}(x)~\partial^{\sigma}u^{\nu}(y) - (\mu \leftrightarrow \nu) ] 
+ \cdots
\eea

\bea
B^{[\mu\nu]\emptyset}_{1}(x,y) = \cdots
+ g_{4}~[\partial^{\mu}\partial_{\rho}\partial_{\sigma} d_{2}(x-y)
~h^{\nu\rho}(x)~u^{\sigma}(y) - (\mu \leftrightarrow \nu) ] + \cdots
\nonumber \\
+ g_{7} [\partial^{\mu}\square d_{2}(x-y)
~u_{\rho}(x)~h^{\nu\rho}(y) - (\mu \leftrightarrow \nu) ] 
\nonumber \\
+ g_{8}~[ \partial^{\mu}\square d_{2}(x-y)
~h^{\nu\rho}(x)~u_{\rho}(y) - (\mu \leftrightarrow \nu) ] + \cdots
\nonumber \\
+ g_{10}~\partial_{\rho}\square d_{2}(x-y)
~h^{\mu\rho}(x)~u^{\nu}(y) - (\mu \leftrightarrow \nu) ] 
\eea

\bea
B^{[\mu][\nu]}_{1}(x,y) = \cdots
+ r_{2}~\partial^{\mu}\partial^{\nu}\partial^{\rho} d_{2}(x-y)~
[ u^{\sigma}(x)~h_{\rho\sigma}(y) + h_{\rho\sigma}(x)~ u^{\sigma}(y) ] 
+ \cdots
\nonumber \\
+ r_{8}~[\partial^{\mu}\square d_{2}(x-y)~u_{\rho}(x)~h^{\nu\rho}(y) 
+ \partial^{\nu}\square~d_{2}(x-y)~h^{\mu\rho}(x)~u_{\rho}(y) ] 
\nonumber \\
+ r_{10}~[ \partial^{\nu}\square d_{2}(x-y) ~u_{\rho}(x)~h^{\mu\rho}(x)
+  \partial^{\mu}\square~d_{2}(x-y)~h^{\nu\rho}(x)~u_{\rho}(y)]
+ \cdots
\nonumber \\
+ r_{16}~\eta^{\mu\nu}~\partial^{\rho}\square d_{2}(x-y)
~[ u^{\sigma}(x)~h_{\rho\sigma}(x) + h_{\rho\sigma}(x)~u^{\sigma}(y)]
\eea

(ii) From the relation
\bea
(\bar{s}B)^{[\mu][\nu]} = D^{[\mu][\nu]} 
\qquad \Leftrightarrow \qquad
\nonumber \\
{\partial \over \partial x^{\rho}}B^{[\mu\rho][\nu]} 
+ (x \leftrightarrow y, \mu \leftrightarrow \nu) - i d_{Q}B^{[\mu][\nu]}
= - i D^{[\mu][\nu]}
\label{coboundary1}
\eea
we obtain many equations; we only select:
\bea
2 a_{1} + 2 a_{3} - 2 a_{9} = - i K_{1}
\nonumber \\
- a_{1} + a_{3} + a_{5} - a_{7} = - i K_{3}
\nonumber \\
- a_{2} + a_{4} - a_{5} - a_{8} = - i K_{4}
\nonumber \\
- 2 a_{3} = - i K_{4}
\nonumber \\
- 2 a_{4} = - i K_{8}
\nonumber \\
2 a_{6} = - i K_{6}
\nonumber\\
- 2 a_{6} = - i K_{2}
\nonumber\\
2 a_{7} + 2 a_{8} + 2 a_{9} = - i K_{5}
\nonumber \\
a_{2} + b_{3} - b_{33} - b_{4} + b_{34} - {1\over 2} r_{2} = i L_{3}
\nonumber\\
a_{4} + b_{4} + b_{46} - {1\over 2} r_{10} = i L_{25}
\nonumber\\
a_{5} - b_{3} - b_{45} - {1\over 2} r_{8} = i L_{9}
\nonumber \\
a_{6} + b_{31} + b_{43} - b_{32} - b_{44} - {1\over 2} r_{16} = i L_{18}
\nonumber \\
- a_{6} + b_{2} + b_{44} - {1\over 2} r_{10} = i L_{24}
\nonumber\\
a_{8} + b_{33} + b_{45} - b_{34} - b_{46} - {1\over 2} r_{16} = i L_{19}
\nonumber \\
b_{1} - b_{31} - b_{2} + b_{32} - {1\over 2} r_{2} = i L_{2}
\nonumber \\
- b_{1} - b_{43} - {1\over 2} r_{8} = i L_{8}
\eea

(iii) From the relation
\bea
(\bar{s}B)^{[\mu\nu]\emptyset} = D^{[\mu\nu]\emptyset} 
\qquad \Leftrightarrow \qquad
\nonumber \\
{\partial \over \partial y^{\rho}}B^{[\mu\nu][\rho]} 
- i d_{Q}B^{[\mu\nu]\emptyset} = - i D^{[\mu\nu]\emptyset}
\label{coboundary2}
\eea
we obtain as above many equations and we select:
\bea
- a_{1} - a_{3} - a_{7} = - i Q_{1}
\nonumber\\
- a_{2} - a_{4} - a_{8} = - i Q_{2}
\nonumber\\
- a_{5} = - i Q_{3}
\nonumber\\
a_{4} - b_{4} - b_{46} + {1\over 2} g_{7} = - i R_{6}
\nonumber\\
- a_{6} - b_{2} - b_{44} + {1\over 2} g_{7} = - i R_{5}
\nonumber\\
- b_{1} - b_{43} - {1\over 2} g_{8} = - i S_{5}
\nonumber\\
- b_{3} - b_{45} - {1\over 2} g_{8} = - i S_{6}
\nonumber \\
- b_{7} - b_{9} - b_{37} - {1\over 2} g_{4} = - i S_{2}
\nonumber \\
- b_{11} - b_{13} - b_{39} - {1\over 2} g_{4} = - i S_{3}
\nonumber \\
- b_{19} - b_{25} = - i S_{7}
\nonumber \\
- b_{21} - b_{27} - {1\over 2} g_{10} = - i S_{9}
\nonumber \\
- b_{22} - b_{29} - {1\over 2} g_{10} = - i S_{8}.
\eea

(iv) Now we can show that the preceding systems are compatible {\it iff} we have
the equations from the statement. It can be proved by direct computations that
no other equations are needed to obtain a solution of the coboundary equation
from the statement. This assertion follows from hard work: one has to write down
the generic expressions for the coboundaries
$
B^{IJ}, |I| + |J| =, 2,3
$
and show that a solution of the equations (\ref{coboundary1}) and
(\ref{coboundary2}) exists {\it iff} the eight equations from the statement are
true.
$\qed$

Now we notice that from the relative cocycle equations we have obtained 
(\ref{(11)}), (\ref{(11.2)}), (\ref{(11.4)}), (\ref{(11.3)}), (\ref{(12.8a)}),
(\ref{(13.8b)}), (\ref{(12.2b)}) and (\ref{(13.5b)}) which are exactly the
equations from the statement of the theorem.
As a conclusion, we have
\begin{cor}
If 
$
D^{IJ}
$
is a cocycle, then we can write
$
D^{IJ}, |I| + |J| = 2
$
as a coboundary.
\end{cor}

\subsection{The Descent Procedure}
We now start a descent procedure. If we use the preceding corollary in the
cocycle 
equation 
\be
(sD)^{[\mu]\emptyset} = 0
\ee
we obtain that the expression
\be
\tilde{D}^{[\mu]\emptyset} \equiv D^{[\mu]\emptyset} 
- i {\partial \over \partial x^{\nu}}B^{[\mu\nu]\emptyset}
+ i {\partial \over \partial y^{\nu}}B^{[\mu][\nu]}
\ee
is a cocycle
\be
d_{Q} \tilde{D}^{[\mu]\emptyset}  = 0.
\ee
Using this cocycle equation one can prove that 
$
\tilde{D}^{[\mu]\emptyset}
$
is in fact a coboundary. For this one must consider all relevant sectors of
this expression. In the sectors 
\bea
\partial\partial\partial\partial d_{2}(x - y)
[ u(x)~h(y) + h(x)~u(y)]
\nonumber \\
\partial\partial\partial\partial d_{2}(x - y)
[ u(x)~\partial h(y) + \partial h(x)~u(y)]
\nonumber \\
\partial\partial\partial\partial d_{2}(x - y)
[ \partial u(x)~\partial h(y) + \partial h(x)~\partial u(y)]
\nonumber
\eea
this result follows elementary. In the sector
\bea
\partial\partial\partial d_{2}(x - y)
[ \partial u(x)~h(y) + h(x)~\partial u(y)]
\nonumber
\eea
we are left with $11$ nontrivial cocycles, some of which cannot be seen
immediatley as coboundaries. For instance
\bea
D^{[\mu]\emptyset}(x, y) = \partial^{\nu}\square d_{2}(x - y)
[ h^{\mu\rho}(x) \partial_{\nu}u_{\rho}(y) 
+ h^{\mu\rho}(x) \partial_{\rho}u_{\nu}(y)
\nonumber \\
+ \partial^{\mu}u^{\rho}(x) h_{\nu\rho}(y) 
+ \partial^{\rho}u^{\mu}(x) h_{\nu\rho}(y)
- \partial^{\rho}u_{\rho}(x) {h^{\mu}}_{\nu}(y)
\nonumber \\
- {1\over 2} \partial_{\nu}u^{\mu}(x) h(y)
- {1\over 2} \partial^{\mu}u_{\nu}(x) h(y)
+ {1\over 2} \delta^{\mu}_{\nu} \partial^{\rho}u_{\rho}(x) h(y) ]
\nonumber
\eea
can be written as
\be
i \partial^{\nu}\square d_{2}(x - y)
d_{Q}[ 2 h^{\mu\rho}(x) h_{\nu\rho}(y) - {h^{\mu}}_{\nu}(x) h(y) ].
\ee

The sector
\bea
\partial d_{2}(x - y)
[ \partial \partial \partial u(x)~\partial h(y) 
+ \partial h(x)~\partial\partial\partial  u(y)]
\nonumber
\eea
is the most complicated one. There are $12$ cocycles in which $\partial h$
appears in the combination
$
\partial_{\sigma}h^{\rho\sigma}
$
so these are seen immediately to be coboundaries. But we are still left with
$9$ nontrivial cocycles as for instance
\bea
D^{[\mu]\emptyset}(x, y) = \partial^{\nu}d_{2}(x - y)
[ \partial_{\nu}h_{\rho\sigma}(x) \partial^{\mu}\partial^{\rho}u^{\sigma}(y) 
+ \partial_{\nu}\partial_{\rho}u_{\sigma}(x) \partial^{\mu}h^{\rho\sigma}(y) ]
\nonumber \\
- \partial^{\mu}d_{2}(x - y)
[ \partial^{\nu}h^{\rho\sigma}(x) \partial_{\mu}\partial_{\rho}u_{\sigma}(y) 
+ \partial_{\nu}\partial_{\rho}u_{\sigma}(x) \partial^{\nu}h^{\rho\sigma}(y) ]
\nonumber
\eea
which can be written as 
\bea
i \partial^{\nu}d_{2}(x - y) d_{Q}
\left[ \partial_{\nu}h_{\rho\sigma}(x) \partial^{\mu}h^{\rho\sigma}(y) 
- {1\over 2}  \partial_{\nu}h(x) \partial^{\mu}(y) \right]
\nonumber \\
- i \partial^{\mu}d_{2}(x - y) d_{Q}
\left[ \partial_{\nu}h_{\rho\sigma}(x) \partial^{\nu}h^{\rho\sigma}(y) 
- {1\over 2}  \partial_{\nu}h(x) \partial^{\nu}(y) \right]
\nonumber
\eea

In the end we prove that
\be
\tilde{D}^{[\mu]\emptyset} = d_{Q}B^{[\mu]\emptyset}
\ee
so we obtain
\be
D^{[\mu]\emptyset} = d_{Q}B^{[\mu]\emptyset}
+ i {\partial \over \partial x^{\nu}}B^{[\mu\nu]\emptyset}
- i {\partial \over \partial y^{\nu}}B^{[\mu][\nu]}
\ee
i.e. the expression
$
D^{[\mu]\emptyset}
$
is a relative coboundary. 

We insert this result in the cocycle equation
\be
(sD)^{\emptyset\emptyset} = 0
\ee
and we obtain that the expression
\be
\tilde{D}^{\empty\emptyset} = D^{\emptyset\emptyset}
- i {\partial \over \partial x^{\nu}}B^{[\mu]\emptyset}
- i {\partial \over \partial y^{\nu}}B^{\emptyset[\mu]}
\ee
is a coboundary
\be
d_{Q}~\tilde{D}^{\empty\emptyset} = 0.
\ee
If we write the generic form of 
$
\tilde{D}^{\emptyset\emptyset}
$
we can prove rather easy that in fact the preceding equation leads to
\be
\tilde{D}^{\emptyset\emptyset} = 0
\ee
so we have
\be
D^{\emptyset\emptyset} =
i {\partial \over \partial x^{\nu}}B^{[\mu]\emptyset}
+ i {\partial \over \partial y^{\nu}}B^{\emptyset[\mu]}
\ee
i.e. a relative cocycle if we take
\be
B^{\emptyset\emptyset} = 0.
\ee
So we have proved the triviality of the cohomology problem. It is important to
stress again that it was not necessary to compute explicitly the expressions
$
D^{IJ}
$.
In the end we have
\begin{thm}
For the pure gravity case, let us consider the expressions
$
D^{IJ}(x,y)
$,
up to the second order of the perturbation theory. Then these are cohomologous
to the 
tree contribution i.e. the loop contribution is trivial. 
\end{thm}

{\bf Proof:} The preceding cohomologous argument has proved the assertion for 
one-loop contributions. For two-loop we have by direct computation the following
non-trivial contribution:
\be
D^{\emptyset\emptyset}_{(2)}(x,y) = i c~\square^{2} d_{3}(x - y)
\ee
(where $c$ is some constant). If we take
\bea
B^{\emptyset\emptyset}_{(2)}(x,y) = 0
\nonumber \\
B^{[\mu]\emptyset}_{(2)}(x,y) = {1\over 2} c~\partial^{\mu}\square d_{3}(x - y)
\eea
then we can write the two-loop contribution as a coboundary.

This means that, up to the second order of the perturbation theory, pure gravity
is a
classical theory. We can consider in the same way the loop contributions coming
from the
interaction between Yang-Mills and gravity. One can prove in fact that the
cocycle equation 
forces this loop contribution to be null.
\newpage
\section{Conclusions}
We have proved that the loop contributions to the causal commutator
$
D^{IJ}_{(1)}
$
are of the form
$
sB +
$
super-renormalizable terms in the Yang-Mills case and simply of the form
$
sB
$
in the pure gravity case. Because the expressions $B$ have also causal support
this property stays true after causal splitting. If
\be
B^{IJ} = B^{IJ,\rm adv} - B^{IJ,ret}
\ee
is a causal splitting, then we have
\be
A^{IJ}_{(1)} = sB^{\rm adv} + {\rm super-renormalizable~terms}
\ee
This means that the main contributions of the perturbation theory are the tree 
contributions which correspond to the classical theory. The quantum corrections 
associated to the loop graphs are behaving better in the ultra-violet limit;
we conjecture that this result stays true in all orders of the perturbation
theory.
So there is a chance to construct a non-perturbative theory for gauge models. 
This follows from the well-known fact that the construction of non-trivial QFT 
models in $1+2$ and $1+1$ dimensions is closely connected to the 
super-renormalizability of the associated perturbation theory. This means that
gauge models are better than say, the
$
\Phi^{4}
$
model in $4$ dimensions, for which it is conjectured that the constructive
quantum field theory does not exists. This is related to the fact that the
$
\Phi^{4}
$
model in $4$ dimensions is only renormalizable (does not have any
super-renormalizable properties for the loop contributions). 

The preceding ideas are a full program for a new line of analysis of quantum
field 
theories. We will continue in another paper with the much modest problem of 
investigating the conjecture in third order of the perturbation theory. 


\newpage
\begin{thebibliography}{99}

\bibitem{BS}
N. N. Bogoliubov, D. Shirkov,
``{\it Introduction to the Theory of Quantized Fields}",
John Wiley and Sons, 1976 (3rd edition)

\bibitem{DF}
M. D\"utsch, K. Fredenhagen,
``{\it A Local (Perturbative) Construction of Observables in Gauge Theories:
the Example of QED}",
Commun. Math. Phys. {\bf 203} (1999) 71-105

\bibitem{DF3}
M. D\"utsch, K. Fredenhagen,
``{\it Algebraic Quantum Field Theory, Perturbation Theory, and the Loop
Expansion}", 
Commun.Math.Phys. {\bf 219} (2001) 5-30

\bibitem{EG}
H. Epstein, V. Glaser,
``{\it The R\^ole of Locality in Perturbation Theory}",
Ann. Inst. H. Poincar\'e {\bf 19 A} (1973) 211-295

\bibitem{Gl}
V. Glaser,
``{\it Electrodynamique Quantique}",
L'enseignement du 3e cycle de la physique en Suisse Romande (CICP), Semestre
d'hiver 1972/73

\bibitem{YM} D. R. Grigore
``{\it On the Uniqueness of the Non-Abelian Gauge Theories in Epstein-Glaser 
Approach to Renormalisation Theory}", 
Romanian J. Phys. {\bf 44} (1999) 853-913

\bibitem{standard} D. R. Grigore
``{\it The Standard Model and its Generalisations in Epstein-Glaser 
Approach to Renormalisation Theory}", 
Journ. Phys. {\bf A 33} (2000) 8443-8476 

\bibitem{fermi} D. R. Grigore
``{\it The Standard Model and its Generalisations in Epstein-Glaser 
Approach to Renormalisation Theory II: the Fermion Sector and the Axial
Anomaly}", 
Journ. Phys {\bf A 34} (2001) 5429-5462

\bibitem{cohomology}
D. R. Grigore, 
``{\it Cohomological Aspects of Gauge Invariance in the Causal Approach}",
Romanian Journ. Phys. {\bf 55} (2010) 386-438

\bibitem{cohomology2}
D. R. Grigore, 
``{\it Perturbative Gravity in the Causal Approach}",
Classical Quant. Gravity {\bf 27} (2010) 015013 (33p)

\bibitem{PS}
G. Popineau, R. Stora, 
``{\it A Pedagogical Remark on the Main Theorem of Perturbative Renormalization
Theory}", unpublished preprint

\bibitem{Sc1}
G. Scharf,
``{\it Finite Quantum Electrodynamics: The Causal Approach}",
(second edition) Springer, 1995

\bibitem{Sc2}
G. Scharf,
``{\it Quantum Gauge Theories. A True Ghost Story}",
John Wiley, 2001
and ``{\it Quantum Gauge Theories - Spin One and Two}",
Google books, 2010

\bibitem{Sto1}
R. Stora,
``{\it Lagrangian Field Theory}",
Les Houches lectures, Gordon and Breach, N.Y., 1971, 
C. De Witt, C. Itzykson eds.

\bibitem{St1}
O. Steinmann,
``{\it Perturbation Expansions in Axiomatic Field Theory}",
Lect. Notes in Phys. {\bf 11}, Springer, 1971

\end{thebibliography}
\end{document}